\documentclass[aps,prd,
superscriptaddress,groupedaddress,nofootinbib]{revtex4} 
\usepackage{amsmath,amssymb}
\usepackage{epsfig,graphicx}
\usepackage{amssymb}
\usepackage{epsfig,color}
\usepackage{pstricks,graphicx,epsfig,color,amssymb,amsmath,amscd}
\usepackage[]{graphicx}

\usepackage{makeidx}
\newcommand{\bi}{\bibitem}
\newcommand{\be}{\begin{eqnarray}}
\newcommand{\ee}{\end{eqnarray}}

\usepackage[]{caption}
\def\-g{\sqrt{-g}}
\renewcommand\rho{\varrho}

\usepackage{multirow}

\begin{document}

\title{\bf Massive primordial black holes in contemporary universe
}
\author{A.~D.~Dolgov}
\email{dolgov@fe.infn.it}
\affiliation{	Novosibirsk State University \\
	Pirogova ul., 2, 630090 Novosibirsk, Russia  }
\affiliation{ Institute of Theoretical and Experimental Physics \\
	Bol. Cheremushkinsaya ul., 25,  113259 Moscow, Russia}%

\author{S.~Porey}
\email{shiladitya@g.nsu.ru}
\affiliation{	Novosibirsk State University \\
	Pirogova ul., 2, 630090 Novosibirsk, Russia  }

\date{\today}


\begin{abstract}
The parameters of the original log-normal mass spectrum of primordial black holes (PBH) are approximately 
adjusted on the basis of existing observational data on supermassive black holes in the galactic centers and 
the mass distribution of the near-solar mass black holes  in the Galaxy. Together with the assumption that PBHs 
make all or a noticeable mass fraction of the cosmological dark matter,  it allows to fix the parameters of the 
original mass spectrum. The predicted, in this way, the number density of MACHOs is found to be about an order 
of magnitude below the observed value. A possible resolution of this controversy may be prescribed to  the 
non-isotropic and inhomogeneous distribution of MACHOs or to the modification of the original spectrum, e.g.
assuming a superposition of two-maximum log-normal spectra  of PBHs. A competing possibility is that MACHOs
are not PBHs but dead primordial compact stars. 
\end{abstract}

\maketitle

\section{Introduction \label{s-intro}}

The idea of the primordial black hole (PBH) formation was first put forward by Zeldovich and Novikov~\cite{ZN1,ZN2} 
and later was elaborated by Carr and Hawking~\cite{hawking-BH,carr-PBH1,carr-PBH2}. According to their ideas, the 
density excess in the early universe might accidentally happen
to be large, $\delta \rho /\rho \sim 1$ at the cosmological horizon scale, and then
that piece of volume would be inside its gravitational radius i.e. it became  a PBH, which
decoupled  from the cosmological expansion. 
 In a subsequent paper Chapline~\cite{chapline} suggested that PBH with masses below the solar mass
might be abundant in the present-day universe with the 
density comparable to the density of dark matter\footnote{We thank P. Frampton for indicating this reference}.

A different mechanism of PBH formation was suggested  in refs.~\cite{Dolgov:1992pu,Dolgov:2008wu}.  
A quarter of century-old idea of these works that  massive and very massive
primordial black holes are abundant in the present-day universe, is gaining more and more popularity. 
According to the mechanism of PBH production proposed in refs.~\cite{Dolgov:1992pu,Dolgov:2008wu}, 
the mass spectrum of PBHs at the moment of creation had the simple log-normal form:
\be
\frac{dN}{dM} = \mu^2 \exp \left[- \gamma \ln^2\left( \frac{M}{M_m} \right) \right]  ,
 \label{dN-dM}
\ee
where $\gamma$ is dimensionless constant and parameters $\mu$ and $M_m$ have the dimension of mass or, what is the same, 
of inverse length (here the natural system of units with $c=k=\hbar =1$ is used). Probably log-normal spectrum is a general feature 
of inflationary production of PBH or, to be more precise, is a consequence of the creation of appropriate conditions for the PBH 
formation during the inflationary cosmological stage, while the PBHs themselves might be formed long after 
inflation. In the mentioned   above
model,  black holes were formed after the QCD phase transition (QCD PT) at the temperature of about 100 MeV.
Some other forms of the spectrum were postulated
in the literature, in particular,  the delta-function one and a power-law spectrum. In this work, we confine ourselves to the log-normal
spectrum which has a rigorous theoretical justification. Such spectrum is an example of the so-called extended mass 
spectrum which has come to life recently, see  e.g. ref.~\cite{clesse-garcia-bellido}, 
instead of narrow (monochromatic) mass spectra assumed in earlier works.
As it was envisaged in ref.~\cite{Dolgov:1992pu}, cosmological dark matter could consist entirely of PBHs 
 (see also ref.~\cite{nov-iv}). 
It was even claimed recently that practically all
black holes in the contemporary universe, with masses starting from a fraction of the solar mass, $M_\odot$, 
up to supermassive black holes of billions  of solar masses, and intermediate-mass black holes with 
$M = (10^3-10^6) \,M_{\odot}$ are predominantly primordial~\cite{beasts,Dolgov:2017aec,Dolgov:2018urv}.
A recent review on the bounds on the contemporary density of PBHs for different masses can be found 
in refs.~\cite{BH-rev,BH-rev2}.

The analysis of the chirp mass distribution of the coalescing BH binaries~\cite{BH-bin-chirp}  agreed very well
with the log-normal mass distribution of the individual black holes and nicely fits the hypothesis that they are
primordial.

It is argued in the paper~\cite{AD-KP-midmass} that the PBH mass distribution has maximum near $M_{BH} = 10 M_\odot$, 
because it is the mass inside the cosmological horizon at the QCD PT.

The recent analysis of the situation with PBH based on the LIGO/Virgo data was performed in ref.~\cite{PBH-LIGO}.
According to the author's conclusion, the current data in the absence of accretion seems to exclude that all the sources of
the observed gravitational waves are primordial. However, the possible effects of accretion can relax this constraint.

In this work, we will use available observational data to fix the parameters of distribution~(\ref{dN-dM}).
This task is highly non-trivial and the results can be trusted only approximately
because the original  mass spectrum of PBHs was surely distorted 
through the matter accretion by PBHs during the course of the cosmological evolution.
This problem was addressed in two works by one of us (with collaborators)~\cite{Blinnikov:2014nea,Dolgov:2017nmh}. 
Here we use a different set of observational data and somewhat 
change the assumptions about the evolution of the original mass  spectrum.

In ref.~\cite{bdpp} we assumed that: \\
1. MACHOs are primordial black holes and their
cosmological mass density makes the fraction $f=0.1$ of the cosmological mass density of dark matter.\\
2. All primordial black holes constitute the whole cosmological dark matter,\\
3. The number density of the primordial black holes  with masses above $10^3 M_\odot$ is equal to the number
density of the observed large galaxies.

The basic assumptions which are relied upon in this work are the following:\\
1. The total cosmological mass density of primordial black holes in the universe makes 
the fraction $f$ of the dark matter density with
$f$ being a free parameter. The most interesting case is of course  $f=1$. We also consider the case $f=0.1$.\\
2. The observed number density of large galaxies is equal to the number density of the heavy black holes with masses exceeding
some boundary value, $M_b$. While $M_b$ is supposed to be much smaller than the masses of the supermassive
black holes (SMBH) observed in  the centers of large galaxies, they could serve as appropriate seeds for the SMBH creation
not only in the present-day universe but also in the young universe at the redshifts $z\sim 10$~\cite{bdpp,AD-KP-midmass,silk-seeds}
\\
3. The value of $M_m$, at which the distribution (\ref{dN-dM}) reaches the maximum, is taken from the data on the mass spectrum 
of black holes in the Galaxy. 
In the papers~\cite{Blinnikov:2014nea,Dolgov:2017nmh},  it was taken to be equal to one solar mass. 
However, in this work, we assume that $M_m$ is in the interval $(6-8) M_\odot $ as dictated by
the observations of the mass spectrum of the black holes in the Galaxy, see Section III. 
Maybe it is more proper to take $M_m$ higher, closer to $10 M_\odot$, as it is argued in ref.~\cite{AD-KP-midmass}.
\\ 
With this choice of the 
three basic sets of the observational data
the mass density of MACHOs derived here is about $f \lesssim 10^{-3}$. The apparent contradiction of the
observations can be resolved if the MACHOs are non-homogeneously distributed in space, see discussion below,
or the mass spectrum (\ref{dN-dM}) is generalized to a more complicated form having two or several maxima, as is 
envisaged in ref.~\cite{Dolgov:2008wu}. Another possibility is that MACHOs  are not PBH but compact primordial
stars which were mostly destroyed after formation~\cite{AD-KP-midmass}.

\section{Total mass density of black holes \label{ss-total-rho}}

The total cosmological mass density of the primordial black holes at the present  time is  given by the integral
\be
\rho_{BH} = \mu^2 \int_0^{M_{max} }dM M  \exp \left[- \gamma \ln^2\left( \frac{M}{M_m} \right) \right] 
\label{rho-bh}
\ee
under the assumption that the spectrum (\ref{dN-dM}) is weakly  distorted by accretion in the  essential mass range
where $M$ is close to $M_m$. 
As shown in ref.~\cite{Blinnikov:2014nea}, the spectrum has a cutoff at large mass, $M_{max}$. 
The maximum value of PBH mass is estimated in  ref.~\cite{Blinnikov:2014nea} as a function of the
model parameters. 
 According to this work, a reasonable value of $M_{max}$ may lay in the range
$M_{max}= (10^5-10^6) M_\odot$. Since $M_m$ is below $10 M_\odot$, see the next section,
integral (\ref{rho-bh}) can be safely extended to infinity.

Assuming that $\rho_{BH} $ makes a fraction $f$ of the mass density of dark matter,  
 $ \rho_{BH}/ \rho_{DM} = f$, where
\be
\rho_{DM} \approx 2.5\ 10^{-30} \,{\rm g/cm}^3 \approx 3.7 \cdot 10^{10} M_\odot/ \rm{Mpc}^3    
\label{rho-DM}
\ee
we find the first equation for the fixation of the parameters of the distribution:
\be
 \mu^2 \int_0^{ M_{max}} dM M  \exp \left[- \gamma \ln^2\left( \frac{M}{M_m} \right) \right] = f\rho_{DM}.
 \label{eq-1}
 \ee

For the numerical estimates, it is convenient to present the solar mass in different units, not only in grams but 
in inverse megaparsec as well:
\be
M_\odot = 2 \cdot 10^{33} \,{\rm g} = 1.75\cdot 10^{95} /{\rm Mpc} .
\label{M-odot}
\ee

There is no agreement on the value of $f$ in the literature. According to the recent work~\cite{Manshanden:2018tze}, the mass fraction 
of black holes should be rather small, $f<  0.1$. However, this result is valid for a high value of the median mass $M_m \geq 20 M_\odot$.
On the other hand, the data on the mass spectrum of Galactic black holes indicate 
that $M_m = (6-9) M_\odot$, see the next section.
For $M_m$ in this interval,
the limits are much weaker. In what follows, we  allow for the extreme case $f = 1$, which  might not be excluded.

\section{Mass spectrum of black holes in the Galaxy \label{BH-gal}}

The mass spectrum of black holes in the Galaxy shows striking features which are not expected in 
the standard picture of stellar-mass BH
formation through stellar collapse after a star exhausted its nuclear fuel and if it had a sufficiently large mass.
The observed picture strongly disagrees with the natural expectation from this scenario. According to ref.~\cite{Ozel:2010su}
the masses of the observed black holes are surprisingly high and are concentrated in a narrow interval $(7.8\pm 1.2) M_\odot$.
This result is supported by another work~\cite{Kreidberg:2012ud}, according to which the spectrum maximum is situated at 
$M \sim 8 M_\odot$ and sharply drops above $M\sim 10 M_\odot $ and below $5 M_\odot$.

It is also observed~\cite{Farr:2010tu} that black holes in the Galaxy have a two-peak mass distribution with the second peak
situated above the maximum mass of neutron star  but below the lower limit of the BH masses found in the quoted above 
papers~\cite{Ozel:2010su,Kreidberg:2012ud}. The lower mass BHs are presumably produced by the usual mechanism of 
stellar collapse. So we expect that galactic black holes have log-normal distribution of heavier BHs, but lower mass BHs 
have a replica of stellar mass distribution of stars exceeding the Chandrasekhar limit.

Matter accretion in the course of galactic evolution may lead to some increase of the galactic black hole masses.
Bearing this in mind, we take as the test values $ M_m/M_\odot = 6,7$, and 8.

\section{Supermassive PBH in the centers of large galaxies \label{s-SM-PBH}}

Astronomical observations strongly indicate that in each large galaxy there resides a supermassive black hole 
(SMBH)~\cite{An-Cher}. 
Moreover,  SMBHs are also observed in some small galaxies and even in practically empty-space;
for a review, see~\cite{Dolgov:2017aec,Dolgov:2018urv}.

The origin of such black holes is mysterious. According to conventional understanding, SMBHs in galactic centers 
appeared as a result of matter accretion on a massive seed. However, the estimates of the necessary accretion rate 
to create such giants demand it to be much larger than any reasonable value. These facts create serious doubts about the 
traditional picture of the galaxy and SMBH formation, according to which the galaxy was created first and later
an SMBH was formed in  the center by the accumulation of the galactic matter.  
The data certainly indicates to invert picture that SMBHs were formed first and  they served as a seed for the galaxy
formation~\cite{Dolgov:1992pu,Dolgov:2008wu,bosch}. 
Recent observations of high red-shift, $z \sim 10$ SMBHs~\cite{Dolgov:2017aec,Dolgov:2018urv}, 
 strongly support this assertion.

Accordingly, we assume that the number density of  supermassive primordial black holes is equal to the 
number density of galaxies.
As it is assumed in ref.~\cite{Blinnikov:2014nea}, the initially formed superheavy PBH might have much smaller masses, 
roughly speaking in the 
range $(10^3 - 10^5)M_\odot$ which could subsequently grow up to $10^9 M_\odot$  because of an efficient accretion
of matter on the preexisting very massive seeds and mergings. 
 A similar statement is done in ref.~\cite{rosas2016supermassive}, namely that the PBHs with masses around  $(10^4 -10^5) M_\odot$
may subsequently grow to $ 10^9 M_\odot$

This mass enhancement factor is much stronger for heavier BH
and thus their mass distribution may be different from (\ref{dN-dM}). We assume the simplified picture that the
original PBHs were created with the distribution (\ref{dN-dM}) but  a PBH with the mass larger than a certain
boundary value $M_b$ became a supermassive seed  for galaxy formation. Correspondingly the number density of
PBH with masses larger than $M_b$ should be equal to the present-day number density of (large) galaxies:
\be
N_b = \mu^2 \int_{M_b}^{M_{max}} dM   \exp \left[- \gamma \ln^2\left( \frac{M}{M_m} \right) \right]  = N_{gal}
\label{N-b}
\ee
In what follows, we take the following two sampling values 
\be
M_b = [ 10^4,\,10^5 ]\,M_\odot .
\label{M-b}
\ee


Evidently, we must choose $M_{max} >  M_b$. If $M_{max} \gg  M_b$, the upper limit in eq.~(\ref{N-b}) maybe extended to infinity.
If accidentally $M_{max}$ is close by magnitude to $M_b$, the integral in Eq. (\ref{N-b}) would be strongly diminished.

The number density of galaxies is not well known. We take it as 
\be
N_{gal} = K /{\rm Mpc}^3 .
\label{N-gal}
\ee
with $K$ presumably in the generous interval $K=( 0.1 - 0.001)$. This estimate is in a reasonable agreement with those presented  
in refs,~\cite{shinkai,monselice}

This relation presents the third and the last necessary condition for the determination of the parameters of distribution (\ref{dN-dM}). 

\section{Determination of the parameters\label{s-param}}

Using the presented above conditions, we can determine the parameters: $\gamma$ and $\mu$. The value of the median
mass $M_m$ is fixed in the interval $6 M_\odot \leq M_m \leq 8 M_\odot$  by the mass spectrum of the Galactic black holes,
see Sec. III.

From equations (\ref{eq-1},\ref{N-b},\ref{N-gal}) we find: 
\be
\frac{\rho_{DM}}{M_{odot} N_{gal}} &=& 3.7\times 10^{10} f /K   \nonumber \\ 
&=& \frac{I_1( 0, x_{max}, x_m, \gamma) }{I_0( x_b, x_{max}, x_m, \gamma))} ,
\label{rho-dm-to-N-gal}
\ee
where

\be
&&I_n(x_{min}, x_{max}, x_m, \gamma) = \nonumber \\
&&\qquad  \int_{x_{min}}^{x_{max}} dx x^n \exp \left[ -\gamma \ln^2 \left(\frac{x}{x_m}\right)\right]
\label{I-n-x}
\ee

with $x_{min} = M_{min}/M_\odot$, $x_{max} = M_{max}/M_\odot$, $x_b = M_b/M_\odot$, and $x_{m} = M_{m}/M_\odot$

We calculate the ratio in the r.h.s. of eq.~(\ref{rho-dm-to-N-gal}) as a function of $\gamma$ for $f=1, 0.1$; $K= 0.1$;
$x_{b} = 10^4, 10^5$ and $x_{max} =  10^5, 10^6$. According to the definition, $M_b$ should be smaller than $M_{max}$ 
in each sample of the parameters. The results are not significantly different except for the case when 
$M_b$ closely approaches  $M_{max}$ from below.

The calculated values of the parameters $\gamma$ and $\mu$ are presented in Table I in the appendix
 for $M_b=10^4, 10^5$. In that table $\mu_1$ is 
the value of parameter $\mu$ calculated from the condition $N_{gal}=0.1/{\rm Mpc}^3$ and $\mu_2$ is the value of the same parameter 
calculated from the condition $\rho_{PBH}= 2.5\ 10^{-30} f \,{\rm g/cm}^3$. As mentioned in Introduction, we have taken two 
sample values of $f$, $1$ and $0.1$.

According to ref.~\cite{Dolgov:2017nmh} fitting the PBH mass function normalization
in the $10-100$~M$_\odot$ range to the BH+BH merging rate derived from the LIGO BH+BH detections  ($ 9-240$
events a year per cubic Gpc), we should only take care that the mass density of primordial SMBHs
does not contradict the existing SMBH mass function as inferred from observations of galaxies,
$dN/(d\log MdV)\simeq 10^{-2}-10^{-3}$~Mpc$^{-3}$. 
The density of  PBHs with the masses in the interval observed by LIGO, i.e. $20 M_\odot - 50 M_\odot$ 
with the chosen values of our parameters, see the left columns of the Table, is equal to a few times $10^9$
per galaxy, if we take the density of galaxies equal to
 $N{gal}= 0.1/{\rm Mpc}^3$. It is a large number, however  such BHs are not all
in a galaxy but dispersed 
in the dark matter halo with the radius an order of magnitude larger than the galactic one. Correspondingly the
number of such BHs in a galaxy would be 2-3 orders of magnitude smaller. 
 Presumably, with so many black holes in disposal, the sufficient number of binaries can be formed to explain 
the observed LIGO rate.

The initially formed superheavy PBHs could  have much smaller masses (around $(10^4 -10^5) M_\odot$
 but still grow up to $ 10^9 M_\odot$ because of an efficient accretion of matter and mergings, 
 see the state-of-the-art SMBH growth calculations in \cite{rosas2016supermassive}.

There is a tremendous activity during several recent years in attempts to derive upper limits on the BH density 
in different mass intervals, see Table II. 
These bounds, however, should be taken with a grain of salt, since the limits are
model-dependent and usually derived with the most favorable assumptions to get the strongest possible bound.


\section{Problems with MACHOs \label{sec-MACHD}}

As we have found in the previous section, $\gamma$ is typically about 0.5. If 
we choose $M_m =(6-8)\,M_\odot$, then the calculated mass density of MACHOs would be 
several orders of magnitude lower than most results on the measured MACHO density for all reasonable values of~$\gamma$. 

The data presented by different groups  are rather controversial. The present date situation is reviewed and summarized in 
refs.~\cite{moniez,sib-ufn,Blinnikov:2014nea,bdpp}. Briefly, the situation is the following. 

MACHO group \cite{MACHO2000} reported the registration of 13 - 17 microlensing events  towards the Large Magellanic Cloud (LMC),
which is significantly higher than the number which could originate from the known low luminosity stars. On the other hand  this amount
is not sufficient to explain all dark matter in the halo. The fraction of the mass density of the observed objects, which created the
microlensing effects, with respect to the energy density of the dark matter in the galactic halo, $f$, according to the 
observations~\cite{MACHO2000} is in the interval:
\be
0.08<f<0.50 ,
\label{f-macho}
\ee
at 95\% CL for the mass range $  0.15M_\odot < M < 0.9M_\odot  $.


EROS 
collaboration~\cite{EROS-1}  has placed the upper
limit on the halo fraction, $f<0.2$ (95\% CL) for { the} objects in the specified above MACHO
mass range, while EROS-2 \cite{Tisserand:2006zx} gives $ f<0.1$ for $0.6 \times 10^{-7}M_\odot<M<15 M_\odot$
for the survey of Large Magellanic Clouds.
It is considerably less than that measured by the MACHO collaboration in the central region of the LMC.

The new  analysis of 2013 
by EROS-2, OGLE-II, and OGLE-III collaborations~\cite{Novati:2013fxa} towards the Small Magellanic Cloud (SMC). 
revealed five  microlensing events towards the SMC (one by EROS and four by OGLE), which lead to the upper limits
 $ {f <0.1} $ obtained at 95\% confidence level for MACHO's with the mass $ 10^{-2} M_\odot$
and $ {f <0.2} $ for MACHOs with the mass $ 0.5 M_\odot$.

Search for microlensing in the direction of Andromeda galaxy (M31) demonstrated some contradicting 
results~\cite{moniez,sib-ufn} with an uncertain conclusion. E.g. AGAPE collaboration \cite{AGAPE2008}, 
finds the halo MACHO fraction in the range $0.2<f<0.9$.
while the MEGA group presented the upper limit $f<0.3$~\cite{MEGA2007}.
On the other hand, the recent discovery of  10 new microlensing events~\cite{Lee-2015}  
is very much in favor of MACHO existence. The authors conclude: ``statistical studies and individual microlensing events
point to a non-negligible MACHO population, though the fraction in the halo mass remains uncertain''.

Some more recent observational data and  the other aspects  of the microlensing are discussed in ref.~\cite{Mao2012}.

It would be exciting if all DM were constituted by old stars and black holes
made from the high-density baryon bubbles 
as suggested in refs.~\cite{Dolgov:1992pu,Dolgov:2008wu} 
with masses in still allowed
intervals,  but a more detailed analysis of this possibility has to be done.

There is a series of papers claiming the end of MACHO era. For example, in ref.~\cite{YooCG2004} the authors 
stated "we exclude MACHOs with masses $M > 43 M_\odot $ at the standard local halo density. This removes 
the last permitted window for a full MACHO halo for masses $ M> 10^{-7.5} M_\odot$.

In addition to the criticism raised in the paper~\cite{YooCG2004}, some more arguments against the abundant
galactic population of MACHOs  are also presented in ref.~\cite{BEDu-2003,BEDu-2004}.
However, according to the paper~\cite{griest},  the approach of the mentioned works have serious
flaws and so their results are questionable. A reply to this criticism is presented in the subsequent 
paper~\cite{EvansBel2007}. 

The data in support  of smaller density of MACHOs in the direction to SMC
is presented in ref.~\cite{Tisserand:2006zx} 

Later, however, another paper of the  Cambridge group \cite{Quinn:2009zg} was published where on the basis of
studies of binary stars, arguments in favor of real existence of MACHOs and against the pessimistic
conclusions of ref.~\cite{YooCG2004} were presented.

The latest investigation on the "end of MACHO era" was presented in ref.~\cite{Monroy-2014}, where it is concluded 
that "the upper bound of the MACHO mass tends to less than $5 M_\odot$
does not differ much from the previous one. Together with microlensing studies that provide lower limits on the MACHO mass, 
our results essentially exclude the existence of such objects in the galactic halo".

A nice review of the state of the art and some new data is
presented in ref.~\cite{Lee-2015} with the conclusion that
some statistical studies and individual microlensing events point to a non-negligible MACHO population, though the fraction 
in the halo mass remains uncertain.

According to the results of different groups, the fraction of MACHO mass density with respect to the total
mass density of dark matter varies in a rather wide range:
\be
f_{MACHO} = \frac{\rho_{MACHO}}{\rho_{DM}} \sim (0.01 - 0.1)
\label{f-MACHO}
\ee
Notice a large variance of the results by different groups. Reasonable agreement between the data and the 
mass spectrum
considered here  can be
achieved only if $ M_m \sim M_\odot $~\cite{Dolgov:2017nmh,bdpp}. So we either have to reject the possibility that practically all
galactic black holes are primordial with masses around $ (6-8) M_\odot$ or to search for another explanation of the discrepancy
between the observed and the predicted density of MACHOs with a log-normal mass spectrum of PBHs.

An interesting option is that the spatial distribution of  MACHOs may be very inhomogeneous and non-isotropic. 
Due to the selection effect, MACHOs are observed
only in over-dense clumps  where their density is much higher than the average one. For a review and
the list of references on dark matter clumping see e.g.~\cite{DM-clump}. Clumping of primordial back holes, 
due to dynamical friction, may be much
stronger than the clumping of dark matter consisting of elementary particles.
This hypothesis would allow to avoid contradiction between the observed high density of MACHOs and the predicted  much smaller density of them based on the log-normal mass spectrum with  $M_m = (7-9) M_\odot $.

Another possibility to adjust theory to the observations is to assume multi-maximum log-normal spectrum i.e.
the superposition of the log-normal spectra with maxima at several different values of $M_m$:
\be
\frac{dN}{dM} = \sum_j \mu^2_j \exp \left[- \gamma_j \ln^2\left( \frac{M}{M_m^j} \right) \right] .
 \label{dN-dM-2}
\ee
Such spectrum may originate from inflationary stage if the coupling of the inflaton field $\chi$ to the scalar with non-zero baryonic 
number has more complicated polynomial form~\cite{Dolgov:2008wu}, than that postulated in the original paper~\cite{Dolgov:1992pu}:
\be
U_{int} = | \chi |^2   \prod_j  \lambda_j (\Phi -\Phi_j)^2/ m_{Pl}^{(2j-2)}.
\label{U-int-2}
\ee 
In our case the two-maxima mass spectrum, with $j$ running from 1 to 2, is sufficient to describe all observational data with
reasonable accuracy. It allows also to avoid many existing bounds on primordial black 
holes~\cite{bound3,bound1,Sasaki:2018dmp,bound4,bound5}.
Such two-maximum log-normal spectrum is introduced ad hoc in ref.~\cite{two-peak} with the same purpose to satisfy the demands
of astronomical  observations.

On the other hand, according to ref.~\cite{AD-KP-midmass} bubbles with high density of baryons  with masses smaller than the 
mass inside cosmological horizon at the QCD phase transition mostly formed compact stellar-like objects, which could  explode in the
process of evolution, leading to a considerably smaller density of MACHOs, than predicted by the log-normal mass spectrum.

\section{Black holes with intermediate mass \label{s-int-mass}}

Black holes with masses from $10^3 M_\odot$ up to $10^6 M_\odot$  are rather arbitrarily called Intermediate Mass Black Holes
(IMBH). They were observed during the recent few years and now about $10^3$ of them are known~\cite{imbh}. It remains unclear
if they can be created by the conventional astrophysical processes, such as stellar collapse or matter accretion to some massive 
seeds. The hypothesis that they are all primordial looks much more attractive.

Having the parameters fixed, we can calculate the number density of the intermediate black holes
 in each galaxy, $N_{IMBH}$. We find that for each large 
galaxy there is $\sim 10^3-10^4$ number of IMBHs (see appendix).
According to ref.~\cite{Dolgov:2017nmh} such IMBH can seed the formation of 
globular clusters and dwarf galaxies. At the moment only in
one globular cluster, a black hole with mass about $2000 M_\odot$ is detected. It is predicted~\cite{Dolgov:2017nmh} that in
every globular cluster there must be an intermediate-mass primordial black hole, which was a seed of this cluster.

\section{Conclusion}

Massive primordial black holes with extended mass spectrum became viable candidates for the constituents of the
cosmological dark matter. Formation of such PBHs is possible due to inflationary expansion of the very early universe because
inflation could create physically connected super-horizon scales. In this sense, the existence of supermassive PBHs can be 
considered as 
extra proof of inflation.

In this work, we relied on the log-normal mass spectrum for the following reasons. First, it was historically first extended 
mass spectrum derived rigorously in a consistent scenario based on a simple modification of the 
Affleck-Dine baryogenesis~\cite{Affleck} 
by the introduction of general renormalizable coupling of the scalar baryon to the inflaton field. Nowadays it is the
very popular form of the mass spectrum of PBHs. 

Moreover, the chirp-mass distribution of the LIGO events analyzed in ref.~\cite{AD-KP-midmass} demonstrates excellent
agreement with log-normal mass distribution of PBHs. 
 
The other reason that we confined ourselves to this particular form of the spectrum is that
the consideration of several different spectra would make the paper too cumbersome to follow. 

Later, there appeared quite a few other models leading to the formation of very massive PBH with extended mass spectrum.
All of them are essentially based on the assumption that the conditions for the PBH formation were prepared during inflation,
as it was pioneered in ref.~\cite{Dolgov:1992pu}. Indeed, if one does not invoke inflation, the PBH mass cannot reach
huge values of $(10^4-10^5) M_\odot$. There are several papers where inflationary PBH formation is studied and extended 
mass spectrum is obtained, see e.g., \cite{Ivanov,Linde-BH,Clesse-BH}.
In ref.~\cite{Ivanov} an extended but quite complicated  mass spectrum was obtained and the possibility of very massive
PBH formation was mentioned. In the paper~\cite{Linde-BH} relatively light PBHs with masses $10^{15} - 10^{30}$~g 
are considered. An extension of the previous work~\cite{Clesse-BH} allows for significantly higher masses  but no clear
statement about the mass spectrum is made. In ref.~\cite{Clesse-BH2} the log-normal mass spectrum was exploited 
but without any justification or reference. The paper~\cite{Carr-BH2} to a large extend 
repeats main features of ref.~\cite{Dolgov:1992pu} 
but with essentially different kinetics.

There are several relatively recent papers dedicated to the consideration of  general conditions for PBH formation 
see e.g.~\cite{Musco,Young}, but the bounds derived there are not applicable to the model of 
refs.~\cite{Dolgov:1992pu,Dolgov:2008wu}, because these general conditions are valid for the traditional mechanism of PBH 
formation, while the mechanism of the works~\cite{Dolgov:1992pu,Dolgov:2008wu} and of some other quoted-above-papers 
are significantly different since PBNs were formed very late, after QCD phase transition and prior to it the 
perturbations were isocurvature ones, which transformed to density perturbations after massless quarks turned into massive
nucleons in bubbles with very high baryonic number.



Recent observations of abundant supermassive black holes
in the early universe lead to a natural conclusion that they are primordial, see e.g.~\cite{Dolgov:2017aec}. If they indeed have
log-normal or some other extended mass distribution, then it is tempting to conclude that the contribution of PBH to the cosmological
dark matter is at least non-negligible. 

In principle, there could be two, or even several, comparable forms of dark matter: PBHs and different elementary particles species, 
though such a conspiracy is surely at odds with the Occam's razor. On the other hand, there are impressive 
examples of similar cosmic 
conspiracies of near equality of energy densities of baryons, dark matter, and dark energy. 
  Detailed comparison of the observational data with the predicted  mass spectrum of black holes at different redshifts could
help to solve this deep mystery.

\section*{Acknowledgements} 
We thank the anonymous referee for many useful comments.

This work was supported by the RNF Grant 19-42-02004.


\newpage
	\section*{Appendix: \label{A-1}}

\begin{table}[h]	
	\begin{center}
		\caption{}
		\begin{tabular}{ |c|c|c|c| } 
			\hline	
			& & $M_b=10^4 M_\odot, M_{max}=10^5 M_\odot$ & $M_b=10^5 M_\odot, M_{max}=10^6 M_\odot$ \\
			\hline
			\hline
			\multirow{22}{*}{$M_m=8M_\odot$}& \multirow{11}{*}{$f=1$}  & $\gamma=0.53$ & $\gamma=0.31$ \\
			& & & \\
			& & $\mu_1=2.4 \times 10^{-50} ({\rm gm \, cm^3})^{-1/2}$ &$\mu_1=8.98 \times 10^{-51} ({\rm gm \, cm^3})^{-1/2}$ \\
			& & $=4.4 \times 10^{-69}{\rm cm}^{-1}$ & $=1.6 \times 10^{-69}{\rm cm}^{-1}$ \\
			& & &  \\
			& & $N_{IMBH}  = 3.4 \times 10^5 $ & $N_{IMBH}= 1.9 \times 10^4 $  \\
			& & (for $10^3\lesssim M_{IMBH}/M_\odot\lesssim 10^4$)& (for $10^4\lesssim M_{IMBH}/M_\odot\lesssim 10^5$)\\
			& & & \\
			& & $\mu_2=2.5 \times 10^{-50} ({\rm gm \, cm^3})^{-1/2}$ &$\mu_2=1.1 \times 10^{-50} ({\rm gm \, cm^3})^{-1/2}$ \\
			& & $=4.6 \times 10^{-69}{\rm cm}^{-1}$ & $=2 \times 10^{-69}{\rm cm}^{-1}$ \\
			& & &  \\
			& & $N_{IMBH}= 3.6 \times 10^5 $ & $N_{IMBH}= 2.8 \times 10^4 $  \\
			& & (for $10^3\lesssim M_{IMBH}/M_\odot\lesssim 10^4$)& (for $10^4\lesssim M_{IMBH}/M_\odot\lesssim 10^5$)\\
			\cline{2-4}
			&  \multirow{11}{*}{$f=0.1$}& $\gamma=0.48$ & $\gamma=0.29$ \\
			& & & \\
			& & $\mu_1=6.4 \times 10^{-51} ({\rm gm \, cm^3})^{-1/2}$ &$\mu_1=3.5 \times 10^{-51} ({\rm gm \, cm^3})^{-1/2}$ \\
			& & $=1.2 \times 10^{-69}{\rm cm}^{-1}$ & $=6.4 \times 10^{-70}{\rm cm}^{-1}$ \\
			& & &  \\
			& & $N_{IMBH}= 8.6 \times 10^4 $ & $N_{IMBH}= 8.6 \times 10^3 $  \\
			& & (for $10^3\lesssim M_{IMBH}/M_\odot\lesssim 10^4$)& (for $10^4\lesssim M_{IMBH}/M_\odot\lesssim 10^5$)\\
			& & & \\
			& & $\mu_2=6.9 \times 10^{-51} ({\rm gm \, cm^3})^{-1/2}$ &$\mu_2=3.1 \times 10^{-51} ({\rm gm \, cm^3})^{-1/2}$ \\
			& & $=1.3 \times 10^{-69}{\rm cm}^{-1}$ & $=5.7 \times 10^{-70}{\rm cm}^{-1}$ \\
			& & &  \\
			& & $N_{IMBH}=  10^5 $ & $N_{IMBH}= 6.8 \times 10^3 $  \\
			& & (for $10^3\lesssim M_{IMBH}/M_\odot\lesssim 10^4$)& (for $10^4\lesssim M_{IMBH}/M_\odot\lesssim 10^5$)\\
			\hline
		\end{tabular}
    
	\end{center}
 \end{table}

	\begin{center}
		\begin{tabular}{ |c|c|c|c| } 
			\hline	
			& & $M_b=10^4 M_\odot , M_{max}=10^5 M_\odot$ & $M_b=10^5 M_\odot, M_{max}=10^6 M_\odot$ \\
			\hline
			\hline
			\multirow{22}{*}{$M_m=7M_\odot$}& \multirow{11}{*}{$f=1$}  & $\gamma=0.51$ & $\gamma=0.31$ \\
			& & & \\
			& & $\mu_1=2.3 \times 10^{-50} ({\rm gm \, cm^3})^{-1/2}$ &$\mu_1=1.3 \times 10^{-50} ({\rm gm \, cm^3})^{-1/2}$ \\
			& & $=4.2 \times 10^{-69}{\rm cm}^{-1}$ & $=2.4 \times 10^{-69}{\rm cm}^{-1}$ \\
			& & &  \\
			& & $N_{IMBH}= 2.6 \times 10^5 $ & $N_{IMBH}= 2.1 \times 10^4 $  \\
			& & (for $10^3\lesssim M_{IMBH}/M_\odot\lesssim 10^4$)& (for $10^4\lesssim M_{IMBH}/M_\odot\lesssim 10^5$)\\
			& & & \\
			& & $\mu_2=2.7 \times 10^{-50} ({\rm gm \, cm^3})^{-1/2}$ &$\mu_2=1.3 \times 10^{-50} ({\rm gm \, cm^3})^{-1/2}$ \\
			& & $=4.9 \times 10^{-69}{\rm cm}^{-1}$ & $=2.4 \times 10^{-69}{\rm cm}^{-1}$ \\
			& & &  \\
			& & $N_{IMBH}= 3.5 \times 10^5 $ & $N_{IMBH}= 2.1 \times 10^4 $  \\
			& & (for $10^3\lesssim M_{IMBH}/M_\odot\lesssim 10^4$)& (for $10^4\lesssim M_{IMBH}/M_\odot\lesssim 10^5$)\\
			\cline{2-4}
			& \multirow{11}{*}{$f=0.1$}& $\gamma=0.47$ & $\gamma=0.28$ \\
			& & & \\
			& & $\mu_1=7.7 \times 10^{-51} ({\rm gm \, cm^3})^{-1/2}$ &$\mu_1=3.2 \times 10^{-51} ({\rm gm \, cm^3})^{-1/2}$ \\
			& & $=1.4 \times 10^{-69}{\rm cm}^{-1}$ & $=5.8 \times 10^{-70}{\rm cm}^{-1}$ \\
			& & &  \\
			& & $N_{IMBH}= 8.5 \times 10^4 $ & $N_{IMBH}= 7.1 \times 10^3 $  \\
			& & (for $10^3\lesssim M_{IMBH}/M_\odot\lesssim 10^4$)& (for $10^4\lesssim M_{IMBH}/M_\odot\lesssim 10^5$)\\
			& & & \\
			& & $\mu_2=7.7 \times 10^{-51} ({\rm gm \, cm^3})^{-1/2}$ &$\mu_2=3.3 \times 10^{-51} ({\rm gm \, cm^3})^{-1/2}$ \\
			& & $=1.4 \times 10^{-69}{\rm cm}^{-1}$ & $=6.0 \times 10^{-70}{\rm cm}^{-1}$ \\
			& & &  \\
			& & $N_{IMBH}= 8.5 \times 10^4 $ & $N_{IMBH}= 7.6 \times 10^3 $  \\
			& & (for $10^3\lesssim M_{IMBH}/M_\odot\lesssim 10^4$)& (for $10^4\lesssim M_{IMBH}/M_\odot\lesssim 10^5$)\\
			\hline
		\end{tabular}
	\end{center}

	\begin{center}
		\begin{tabular}{ |c|c|c|c| } 
			\hline	
			& & $M_b=10^4 M_\odot, M_{max}=10^5 M_\odot$ & $M_b=10^5 M_\odot, M_{max}=10^6 M_\odot$ \\
			\hline
			\hline
			\multirow{22}{*}{$M_m=6M_\odot$}& \multirow{11}{*}{$f=1$}  & $\gamma=0.5$ & $\gamma=0.3$ \\
			& & & \\
			& & $\mu_1=3.2 \times 10^{-50} ({\rm gm \, cm^3})^{-1/2}$ &$\mu_1=1.3 \times 10^{-50} ({\rm gm \, cm^3})^{-1/2}$ \\
			& & $=5.8 \times 10^{-69}{\rm cm}^{-1}$ & $=2.4 \times 10^{-69}{\rm cm}^{-1}$ \\
			& & &  \\
			& & $N_{IMBH}= 2.9 \times 10^5 $ & $N_{IMBH}= 1.9 \times 10^4 $  \\
			& & (for $10^3\lesssim M_{IMBH}/M_\odot\lesssim 10^4$)& (for $10^4\lesssim M_{IMBH}/M_\odot\lesssim 10^5$)\\
			& & & \\
			& & $\mu_2=3.1 \times 10^{-50} ({\rm gm \, cm^3})^{-1/2}$ &$\mu_2=1.4 \times 10^{-50} ({\rm gm \, cm^3})^{-1/2}$ \\
			& & $=5.7 \times 10^{-69}{\rm cm}^{-1}$ & $=2.6 \times 10^{-69}{\rm cm}^{-1}$ \\
			& & &  \\
			& & $N_{IMBH}= 2.7 \times 10^5 $ & $N_{IMBH}= 2.2 \times 10^4 $  \\
			& & (for $10^3\lesssim M_{IMBH}/M_\odot\lesssim 10^4$)& (for $10^4\lesssim M_{IMBH}/M_\odot\lesssim 10^5$)\\
			\cline{2-4}
			& \multirow{11}{*}{$f=0.1$}& $\gamma=0.45$ & $\gamma=0.27$ \\
			& & & \\
			& & $\mu_1=7.5 \times 10^{-51} ({\rm gm \, cm^3})^{-1/2}$ &$\mu_1=3.0 \times 10^{-51} ({\rm gm \, cm^3})^{-1/2}$ \\
			& & $=1.4 \times 10^{-69}{\rm cm}^{-1}$ & $=5.5 \times 10^{-70}{\rm cm}^{-1}$ \\
			& & &  \\
			& & $N_{IMBH}= 6.7 \times 10^4 $ & $N_{IMBH}= 5.9 \times 10^3 $  \\
			& & (for $10^3\lesssim M_{IMBH}/M_\odot\lesssim 10^4$)& (for $10^4\lesssim M_{IMBH}/M_\odot\lesssim 10^5$)\\
			& & & \\
			& & $\mu_2=8.4 \times 10^{-51} ({\rm gm \, cm^3})^{-1/2}$ &$\mu_2=3.5 \times 10^{-51} ({\rm gm \, cm^3})^{-1/2}$ \\
			& & $=1.5 \times 10^{-69}{\rm cm}^{-1}$ & $=6.4 \times 10^{-70}{\rm cm}^{-1}$ \\
			& & &  \\
			& & $N_{IMBH}= 8.3 \times 10^4 $ & $N_{IMBH}= 8 \times 10^3 $  \\
			& & (for $10^3\lesssim M_{IMBH}/M_\odot\lesssim 10^4$)& (for $10^4\lesssim M_{IMBH}/M_\odot\lesssim 10^5$)\\
			\hline
		\end{tabular}
	\end{center}

\begin{table}[ht]
	\centering
		\caption{\label{Table II}}
	\begin{tabular}{c|c|c|c}
		\hline\hline
		Eridanus II star cluster & $(10-100)M_\odot$ & $f_{\rm PBH, max}>0.02$   & Fig.[3] of \cite{Brandt:2016aco} \\
		\hline 
		EROS-2 and MACHO (milky way) & $10^{-2}\leq M/M_\odot\leq 100$  &$f_{\rm PBH}>2\times 10^{-2}$   & Fig.[8] of \cite{Calcino:2018mwh} \\
		\hline
		GW simulations & $(2-160)M_\odot$ & $f_{\rm PBH}\approx 0.002$& \cite{Raidal:2018bbj}\\
		\hline 
		Radio and x-ray emission and simulations & $(1-100)M_\odot$ & $f_{\rm PBH}> 2 \times 10^{-4}$  & Fig.[5] of \cite{Manshanden:2018tze} \\
		\hline 
		OGLE concluding paper & $M<0.1M_\odot$ & $f_{\rm PBH}<4\%$&\cite{Wyrzykowski:2011tr} \\
		 (towards Magellanic clouds)&$(0.1-0.4)M_\odot$ & $f_{\rm PBH}\sim6\%$&  \\
		& $M_\odot$ & $f_{\rm PBH}< 9\%$ &  \\
		& $20M_\odot$ & $f_{\rm PBH}< 20\%$ &  \\
		\hline 
		PBH and LIGO first run conclusion &$0.2M_\odot$ & $f_{\rm PBH}<5\%$ & (LIGO colla.)\cite{Abbott:2018oah}  \\
		(Monochromatic mass distribution)&$M_\odot$ & $f_{\rm PBH}<5\%$&   \\
		\hline 
		New analysis of EROS-2, and OGLE-II, and & $ 10^{-2} < M/M_\odot < 10^{-1}$ & $f_{\rm PBH}\leq0.11-0.13$ & Fig.[6] of \cite{Novati:2013fxa}  \\
		OGLE-III towards SMC &$M/M_\odot=1,10^{-3}$ &$f_{\rm PBH}\leq0.3$  & \\
		& $ 10^{-3} < M/M_\odot < 10^{-1}$ & $f_{\rm PBH}\leq0.10$ & \\
		&$M/M_\odot=10^{-2}$ &$f_{\rm PBH}\leq0.07$  & \\
		&$M/M_\odot=1$ &$f_{\rm PBH}\leq0.35$  & \\
		\hline 
		New analysis of OGLEII and Ogle III & $10^{-2} < M/M_\odot < 0.5$ & $f_{\rm PBH}\leq 0.1-0.2$ & \cite{Novati:2011ii}   \\
		towards LMC & $M/M_\odot=1$ & $f_{\rm PBH}=0.24$ & \\
		& $10^{-2} < M/M_\odot < 0.1$ & $f_{\rm PBH}\leq0.5$ & \\
		& $M/M_\odot=100$ & $f_{\rm PBH}\sim 0.5$ & \\
		\hline 
		Planck data &  $0.2\leq M/M_\odot \leq 100$ & $f_{\rm PBH}>10^{-6}$ & Fig.[4] of \cite{Chen:2016pud}\\
		\hline 
		EROS-2 (towards Magellonic clouds))&  $ 10^{-3} < M/M_\odot < 10^{-1}$ & $f_{\rm PBH} < 0.04 $ & \cite{Tisserand:2006zx}  \\
		&  $10^{-6} < M/M_\odot < 1$ & $f_{\rm PBH} < 0.1$ &  \\
		\hline 
		GW obsevations & $10\leq M/M_\odot \leq 200 $  & $f_{\rm PBH}<0.008$ & Fig.[17] of\cite{Sasaki:2018dmp}  \\
		\hline 
		Wide binary & $20<M/M_\odot\leq 100$ & $f_{\rm PBH}\geq 0.4 $  & Fig.[3] of \cite{Quinn:2009zg}  \\
		\hline 
		CMB limits on accreting PBHs & $10\leq M/M_\odot \leq 100$ & $f_{\rm PBH}\geq 60$ & Fig.[14] of \cite{Ali-Haimoud:2016mbv} \\
		\hline 
		CMB bounds on disk-accreting PBHs& $0.1<M/M_\odot \geq 100$ & $f_{\rm PBH}>6\times 10^{-5}$  & Fig.[4] of \cite{Poulin:2017bwe}  \\
		\hline 
		Radio and X-ray bound & $20< M/M_\odot<100$ &  $f_{\rm PBH}>2 \times 10^{-2}$  & Fig[1] of \cite{Gaggero:2016dpq}  \\
		\hline 
		WMAP & $1\leq M/M_\odot \leq 100$ &  $f_{\rm PBH}>4 \times10^{-3}$  & Fig.[2] of \cite{Inoue:2017csr}  \\
		& & & (original Fig.[9] of \cite{Ricotti:2007au})  \\
		\hline
	\end{tabular}
\end{table}

\end{document}